\documentclass[aps,pra,twocolumn,showpacs,eqsecnum,groupedaddress,floatfix]{revtex4}
\usepackage{graphicx}
\usepackage{bm}

\begin{document}


\title{Non-Markovian thermalization of entangled qubits}
\author{Ian Glendinning}
\affiliation{European Center for Parallel Computing at Vienna
(VCPC), Nordbergstr.\ 15/C/3, A-1090 Vienna, Austria}
\author{Matthias Jakob}
\affiliation{ARC Seibersdorf Research GmbH, Tech Gate Vienna,
Donau-City-Str.\ 1, A-1220 Vienna, Austria}
\author{Michael N\"olle}
\affiliation{ARC Seibersdorf Research GmbH, Tech Gate Vienna,
Donau-City-Str.\ 1, A-1220 Vienna, Austria}
\date{\today}
\begin{abstract}
We study the decoherence properties of an entangled bipartite qubit
system, represented by two two-level atoms that are individually
coupled to non-Markovian reservoirs. This coupling ensures that the
dynamical equations of the atoms can be treated independently. The
non-Markovian reservoirs are described by a model which leads to an
exact non-Markovian master equation of the Nakajima-Zwanzig form [J.
Salo, S\ M.\ Barnett, and S.\ Stenholm, \oc {\bf{259}}, 772 (2006)].
We consider the evolution of the entanglement of a two-atom state
that is initially completely entangled, quantified by its
concurrence. Collapses and revivals in the concurrence, induced by
the memory effects of the reservoir, govern the dynamics of the
entangled quantum system. These collapses and revivals in the
concurrence are a strong manifestation of the non-Markovian
reservoir.
\end{abstract}
\maketitle
\section{Introduction}\label{Sec.1}

The dynamics of open quantum systems is usually considered under the
assumption that the physical systems of interest are sufficiently
isolated from their environment. This allows one to utilize certain
approximations such as the weak coupling and the Markovian
approximation \cite{Breuer}. The first approximation ensures that
the interaction between the system and the environment is
sufficiently weak so that the system is quasi closed. The latter one
assumes that the characteristic times of the reservoir are much
shorter than that of the system. This allows one to ignore memory
effects from the environment and one arrives at time-local equations
of motion for the reduced state of the system which are ususally
expressed in terms of Gorini-Sudarshan-Kossakowski-Lindblad
operators \cite{Lindblad,Gorini}. These operators generate a
completely positive time evolution which conserve the physical
properties of all quantum states \cite{Benatti}.

Recent studies have shown the limits of the Markovian description of
quantum computation and quantum error correction
\cite{Alicki1,Alicki2,Ahn1,Ahn2,Daffer,Terhal,Aliferis}. In quantum
information theory the physical conditions underlying the Markovian
approximation can be violated since the coupling of the reduced
system to the environment may be of the same order of magnitude as
the environment relaxation rate. Consequently, the time evolution of
the reduced system is no longer Markovian and the equations of
motion either become temporally non-local Nakajima-Zwanzig equations
\cite{Nakajima,Zwanzig} or explicitly time dependent
time-convolutionless equations \cite{Shibata}, depending on the
approach chosen \cite{Breuer}. A feature that these equations share
is that they are often difficult to derive for any given model, and
also to solve. Moreover, they may lead to non-physical behavior such
as the violation of positivity of the dynamical map
\cite{Daffer,Barnett1,Budini,Shabani,Maniscalco,Maniscalco2} which
is a consequence of the phenomenological nature of most of the
non-Markovian approaches. Since there are currently no established
criteria that can be used to identify non-Markovian master equations
that preserve complete positivity, the only way to ensure this
property is to find a model that itself generates completely
positive maps and to use it with no approximations in order to
derive the final equations of motion. This has recently been
achieved by Salo, Barnett and Stenholm \cite{Salo} by considering
the problem of non-Markovian thermal damping of a two-level atom.
The model consisted of coupling the principal atom to an additional
two-level atom which in turn is coupled to a thermal environment.
Utilizing the exact Nakajima-Zwanzig elimination procedure enables
them to find a master equation with memory that is guaranteed to
generate completely positive evolution.

\begin{figure}[!ptb]
\centerline{\includegraphics[width=6cm]{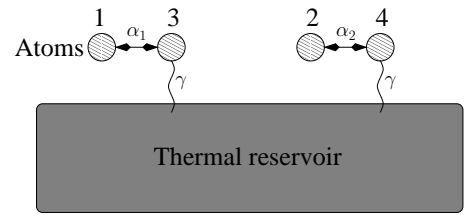}} \caption{Model
for the non-Markovian decay of an entangled two-atom system. Two
(entangled) atoms $1$ and $2$ are individually coupled to two
additional atoms $3$ and $4$, which are in turn coupled to a thermal
reservoir with decay rates $\gamma$. } \label{FigConfig}
\end{figure}

In this paper we discuss the dynamics of the entanglement of a
bipartite system in a non-Markovian reservoir. The system is
composed of two qubits which are represented by two-level atoms. In
order to derive an exact non-Markovian master equation we extend the
model of Salo {\it{et al.}} \cite{Salo} to an entangled two-atom
system to guarantee its completely positive evolution. As in
\cite{Salo}, the two atoms are {\textit{individually}} coupled to
two additional two-level atoms which in turn interact individually
with a thermal environment, as shown in Fig.\ \ref{FigConfig}.
This ensures that the reservoirs are not able to entangle the
two-atom system and, consequently, it is possible to derive
non-Markovian master equations which are of a similar form to the
equations in \cite{Salo}. However, although the reservoir is not
able to entangle the two-atom systems the dynamics of the
entanglement of an initially entangled two-atom system shows a
behavior which clearly demonstrates the memory effects of the
non-Markovian reservoir. In particular, we show that the
entanglement, quantified by the concurrence
\cite{Wootters1,Wootters2}, shows an oscillating behavior in time
that depends on the degree to which the reservoir is non-Markovian.
Thus, the reservoir remembers that the two-atom system was initially
entangled, and the entanglement, which may vanish during the time
evolution, can be retrieved by the system, leading to a ``collapse
and revival like'' time evolution of the entanglement. Such an
behavior may be of importance in quantum computational systems that
are exposed to non-Markovian decoherence effects because even the
concurrence of an initially {\textit{completely}} entangled state
can vanish after a finite period of time \cite{Eberly1,Eberly2}.
Note, however, that the concurrence will partially revive due to the
memory of the reservoir that feeds back into the system. Note
further, that this model allows decoherence effects to be modified
by adjusting the parameters between the interacting atoms. This is
to some extent similar to the modification of decoherence
\cite{Jakob1,Jakob2} in engineered reservoirs
\cite{Monroe1,Monroe2}.

The rest of this paper is organized as follows: In Sec.\ \ref{Sec.2}
we introduce the parameterized thermalization model for the
entangled qubit system. In Sec.\ \ref{Sec.3} we derive the dynamical
equations of the coupled two-atom system and solve them numerically,
assuming that the system is initially in a maximally entangled
quantum state of the Bell form. We discuss the dynamical evolution
of the entanglement which is quantified by the concurrence, in Sec.\
\ref{Sec.4}. We also discuss the effects of a finite-temperature
non-Markovian reservoir on the dynamical evolution of the
concurrence. We conclude in Sec.\ \ref{Sec.Concl} with some
discussion.
\section{Parameterized thermalization model for two entangled qubits}\label{Sec.2}
In this section we derive a simple toy model consisting of two
entangled qubits that are {\textit{individually}} coupled to
non-Markovian thermal reservoirs. This model is based on a recently
proposed system that consists of a single two-level atom coupled to
a specific non-Markovian thermal bath, such that it is possible to
derive an exact master equation that preserves complete positivity
\cite{Salo}. The system is composed of a two-level atom $1$ that is
coupled via an exchange interaction to another atom $2$, which, in
turn, is damped by a thermal reservoir at a specific temperature
that can be characterized with the bosonic excitation number
$\overline{n}$. We extend this model to an entangled two-atom system
$1$ and $2$ where both atoms are assumed to be two-level atoms which
are {\textit{individually}} coupled to two additional and possibly
different two-level atoms $3$ and $4$ via an exchange interaction.
The atoms $3$ and $4$ are both damped by a thermal reservoir at a
temperature that is characterized with the bosonic excitation number
$\overline{n}$. Consequently, the atoms $3$ and $4$ play the role of
a ``near-environment'' to the atoms $1$ and $2$, respectively. Their
dynamical influence produces memory effects in the subsystem of atom
$1$ and $2$. The individual coupling of the atoms to the reservoir
ensures that the dynamics of the two atoms can be treated
independently. We note, that in the single-atom case at zero
temperature with no initial environment photons present, this system
corresponds to the damped Jaynes-Cummings model \cite{Barnett}, as
discussed in \cite{Salo}.

Suppose the atoms $1$ and $2$ are initially in an arbitrary state.
Further, assume that the atoms $3$ and $4$ are initially
uncorrelated with the atoms $1$ and $2$ and that they are both in a
thermal state
\begin{eqnarray}
\overline{\rho}_{3}&=& \overline{\rho}_{4}
=\overline{\rho}\equiv\left[
\begin{array}{cc}
\frac{\overline{n}}{2\overline{n}+1} & 0 \\
0 & \frac{\overline{n}+1}{2\overline{n}+1}
\end{array}
\right]. \label{eq2.1}
\end{eqnarray}
The initial state of the total system is thus described as
\begin{equation}
\rho(t=0)\equiv\rho_{1234}(t=0)=
\rho_{12}(t=0)\otimes\overline{\rho}_{3}\otimes\overline{\rho}_{4}.
\label{eq2.3}
\end{equation}
Here, $\rho_{12}$ is an arbitrary density operator of the subsystem
$1-2$ that allows us to describe initially entangled states of atoms
$1$ and $2$. The atomic states represent the computational qubit
basis $|0\rangle$ and $|1\rangle$ for the atomic ground and excited
state, respectively.

The dynamics of the composite four-atom system is governed by the
following equation ($\hbar\equiv 1$)
\begin{eqnarray}
\frac{d\rho(t)}{dt} &=&
-i\left[\sum_{k=1}^{4}H_{k}+H_{\text{int}}^{(13)}+
H_{\text{int}}^{(24)},\rho(t)\right] \nonumber
\\
&& +{\cal{L}}_{3}\rho(t)+{\cal{L}}_{4}\rho(t) \nonumber \\
&\equiv & {\cal{L}}\rho(t)=
[{\cal{L}}^{(1)}+{\cal{L}}^{(2)}]\rho(t), \label{eq2.4}
\end{eqnarray}
where,
\begin{eqnarray}
{\cal{L}}^{(1)} &=& -i
\left[H_{1}+H_{3}+H_{\text{int}}^{(13)},\rho(t)\right]+{\cal{L}}_{3}\rho(t),
\label{eq2.5aa} \\
{\cal{L}}^{(2)} &=& -i
\left[H_{2}+H_{4}+H_{\text{int}}^{(24)},\rho(t)\right]+{\cal{L}}_{4}\rho(t),
\label{eq2.5ab} \\
H_{n} &=& \omega_{n} |1\rangle_{nn}\langle 1|,\qquad n=1,2,3,4 \label{eq2.5a} \\
H_{\text{int}}^{(13)} &=&
\alpha_{1}\left(\sigma_{1}^{+}\sigma_{3}^{-}+\sigma_{1}^{-}\sigma_{3}^{+}\right),
\label{eq2.5e} \\
H_{\text{int}}^{(24)} &=&
\alpha_{2}\left(\sigma_{2}^{+}\sigma_{4}^{-}+\sigma_{2}^{-}\sigma_{4}^{+}\right),
\label{eq2.5f} \\
{\cal{L}}_{3}\rho &=& \gamma(\overline{n}+1)
(2\sigma_{3}^{-}\rho\sigma_{3}^{+}-\sigma_{3}^{+}\sigma_{3}^{-}\rho-\rho\sigma_{3}^{+}\sigma_{3}^{-})
\nonumber \\
&& +\gamma\overline{n}
(2\sigma_{3}^{+}\rho\sigma_{3}^{-}-\sigma_{3}^{-}\sigma_{3}^{+}\rho-\rho\sigma_{3}^{-}\sigma_{3}^{+}),
\label{eq2.5g} \\
{\cal{L}}_{4}\rho &=& \gamma(\overline{n}+1)
(2\sigma_{4}^{-}\rho\sigma_{4}^{+}-\sigma_{4}^{+}\sigma_{4}^{-}\rho-\rho\sigma_{4}^{+}\sigma_{4}^{-})
\nonumber \\
&& +\gamma\overline{n}
(2\sigma_{4}^{+}\rho\sigma_{4}^{-}-\sigma_{4}^{-}\sigma_{4}^{+}\rho-\rho\sigma_{4}^{-}\sigma_{4}^{+}).
\label{eq2.5h}
\end{eqnarray}
Here, $\omega_{n}$, $n=1,2,3,4$ is the energy of the excited state
of atom $n$, $\alpha_{1}$ and $\alpha_{2}$ are the coupling
strengths between atoms $1-3$ and $2-4$, respectively, $\gamma$ is
the thermalization rate for atoms $3$ and $4$, and $\overline{n}$ is
the thermal bosonic excitation number. In order to simplify matters
we have assumed that the atoms $3$ and $4$ are coupled to a single
reservoir, although in principle they could be coupled to separate
reservoirs with different temperatures. Note, that the dynamical
equation (\ref{eq2.4}) is unable to generate entanglement between
the atoms $1$ and $2$ since the equations of motion decouple between
the atomic subsystems $1-3$ and $2-4$. However, if entanglement is
initially present in the atomic subsystem $1-2$, the memory effects
which emerge from the coupling to the atomic subsystems $3$ and $4$
will be able to modify the decoherence rate of the entanglement.
Consequently, the non-Markovian bath dynamically influences the
decoherence rate of the entanglement of subsystem $1-2$ because of
the feedback of the atomic subsystems $3$ and $4$ on the system
$1-2$ under consideration.

The underlying idea of this non-Markovian bath model is to assume
the atoms $3$ and $4$ serve as ``mini-reservoirs'' which are able to
dynamically modify the equations of motion of the relevant atoms $1$
and $2$. The coupling between the mini-reservoirs can be steered via
the interaction parameters $\alpha_{1}$ and $\alpha_{2}$ as well as
the detuning parameters $\Delta_{k}=w_{k}-\omega_{k+2}$ for $k=1,2$.
These mini-reservoirs will be unobserved and because of their
couplings to a thermal Markovian reservoir they are assumed to
remain in a certain reference state, namely a thermal reservoir
state of a temperature that is imposed by the thermal environment.
Their dynamical influence which feeds back into the subsystem $1-2$,
however, introduces non-Markovian effects in the atomic subsystem
$1-2$. This remains true even in the case when they remain
unobserved and are assumed to be in the thermal reference state, as
it has been demonstrated in \cite{Salo}.

The Nakajima-Zwanzig equation of motion \cite{Nakajima,Zwanzig}
projects the combined quantum system into the {\textit{relevant}}
parts, described by atom $1$ and $2$, and the {\textit{irrelevant}}
parts, represented by atoms $3$ and $4$, utilizing projector
operators ${\cal{P}}_{k}$ and ${\cal{Q}}_{k}$, where $k=1,2$
indicate the atomic subsystems $1$ and $2$. When we insert the
thermal state as the reference state for the atoms $3$ and $4$, the
projectors are given as
\begin{eqnarray}
{\cal{P}}_{1}\rho_{13} &=&
\text{Tr}_{3}[\rho_{13}]\otimes\overline{\rho}_{3}, \label{eq2.6a} \\
{\cal{P}}_{2}\rho_{24} &=&
\text{Tr}_{4}[\rho_{24}]\otimes\overline{\rho}_{4}, \label{eq2.6b} \\
{\cal{Q}}_{1} \rho_{13}&=&
\rho_{13}-\text{Tr}_{3}[\rho_{13}]\otimes\overline{\rho}_{3},
\label{eq2.6c} \\
{\cal{Q}}_{2}\rho_{24} &=&
\rho_{24}-\text{Tr}_{4}[\rho_{24}]\otimes\overline{\rho}_{4},
\label{eq2.6d}
\end{eqnarray}
where $\rho_{13}$ and $\rho_{24}$ define arbitrary operators (not
necessarily density operators) of the atoms $1-3$ and $2-4$,
respectively. Note, that in general, the projectors can operate on
arbitrary operators of the combined systems $1-3$ and $2-4$. The
Nakajima-Zwanzig equation of motion has been derived for the single
atom case in \cite{Salo} under the assumption that, initially, the
atom is not entangled with the ``mini-reservoir atom'' at $t=0$ and
the mini-reservoir atom is in the thermal state $\overline{\rho}$.
In the two-atom case, the situation which is discussed in this
paper, the Nakajima-Zwanzig equations of motion turn out to be of
the same form,
\begin{eqnarray}
\frac{d}{dt}{\cal{P}}_{k} \rho^{(k)}(t)&=&
{\cal{P}}_{k}{\cal{L}}^{(k)}{\cal{P}}_{k}\rho^{(k)}(t) \nonumber
\\
&&\!\!\!\!\!\!\!\! + \int_{0}^{t}
{\cal{P}}_{k}{\cal{L}}^{(k)}{\cal{G}}^{(k)}(t-\tau){\cal{Q}}_{k}{\cal{L}}^{(k)}
{\cal{P}}_{k}\rho^{(k)}(\tau)d \tau .
\nonumber \\
&& \label{eq2.7a}
\end{eqnarray}
This follows from the fact that the dynamical equation of motion
(\ref{eq2.4}) is not able to entangle the atoms $1$ and $2$.
Consequently, we can rewrite every density operator of an entangled
state between atoms 1 and 2 as a linear combination of tensor
products of operators on systems 1 and 2 for which the single atom
Nakajima-Zwanzig formalism of \cite{Salo} applies. Here,
${\cal{G}}^{(k)}(t-\tau)=\exp[(t-\tau){\cal{Q}}_{k}{\cal{L}}^{(k)}]$
where it is assumed that ${\cal{L}}^{(k)}$ is time-independent and
$\rho^{(k)}(t)$, $k=1,2$ represents the operators $\rho_{13}$ and
$\rho_{24}$, respectively.

The Nakajima-Zwanzig equation of motion has been derived in
\cite{Salo} to which we refer for further details. What is important
in these master equations is that the operator expressions
describing the memory terms are not of the Lindblad form. In
addition, the memory kernel contains two different memory functions
one for the diagonal part of the density operators and another one
for the coherences. Hence the total convolution kernel can not be of
the form $K(t-\tau){\cal{L}}$ where ${\cal{L}}$ is a Lindblad
operator. Nevertheless, under the assumption (\ref{eq2.3}) of the
initial state of the total combined system the Nakajima-Zwanzig
equation of motions is an exact equation that can be derived without
any approximations and its solutions remain legitimate density
operators at all times. In other words the dynamical equation of the
subsystem $1-2$ is described by a completely positive map. In the
next section we solve the master equation (\ref{eq2.4}) by
projecting the total density operator onto the subspace in which the
atoms 3 and 4 remain in the thermal reference state given in Eq.\
(\ref{eq2.1}) during time evolution. This ensures that the density
operator $\rho_{12}(t)$ of the subsystem $1-2$ is a legitimate
density operator at all times.

\section{Dynamical equations of the composite two-atom
system}\label{Sec.3}
In this section we numerically solve the dynamical equations of the
density operator $\rho_{12}(t)$ of an initially entangled two-atom
system in the non-Markovian reservoir which is represented by the
model of the preceding section. Initially, we suppose the two atom
system to be in a completely entangled Bell state,
\begin{equation}
|\Psi\rangle_{12}(t=0) =
\frac{1}{\sqrt{2}}\left[|0\rangle_{1}\otimes|0\rangle_{2}+
|1\rangle_{1}\otimes|1\rangle_{2} \right]. \label{eq3.1}
\end{equation}
The corresponding density matrix is accordingly given as
\begin{eqnarray}
\rho_{12}(t=0) &=& \left[
\begin{array}{cccc}
1/2 & 0 & 0 & 1/2 \\
0 & 0 & 0 & 0 \\
0 & 0 & 0 & 0 \\
1/2 & 0 & 0 & 1/2
\end{array}
\right] \nonumber \\
&=&
 \frac{1}{2}\left[\rho_{00}^{(1)}\otimes\rho_{00}^{(2)} +
\rho_{01}^{(1)}\otimes\rho_{01}^{(2)} \right. \nonumber \\
&& + \left. \rho_{10}^{(1)}\otimes\rho_{10}^{(2)}
+\rho_{11}^{(1)}\otimes\rho_{11}^{(2)}\right], \label{eq3.2}
\end{eqnarray}
where the elements are given as
\begin{eqnarray}
\rho_{00}^{(k)}(t=0) &=& \left[
\begin{array}{cc}
0 & 0 \\
0 & 1
\end{array}
\right], \qquad \rho_{11}^{(k)}(t=0) =\left[
\begin{array}{cc}
1 & 0 \\
0 & 0
\end{array}
\right], \nonumber \\
&& \label{eq3.3}
\\
\rho_{10}^{(k)}(t=0) &=& \left[
\begin{array}{cc}
0 & 1 \\
0 & 0
\end{array}
\right], \qquad \rho_{01}^{(k)}(t=0) = \left[
\begin{array}{cc}
0 & 0 \\
1 & 0
\end{array}
\right], \nonumber
\end{eqnarray}
for all $k \in 1,2$. We represent the density matrix of the total
system $\rho_{1234}(t)$ using the operator basis derived in
\cite{Salo}. In this basis we can express
\begin{equation}
\rho_{1234}(t) =
\sum_{m,n=0}^{8}c_{m}^{(1)}(t)c_{n}^{(2)}(t)X_{m}^{(1)}\otimes
X_{n}^{(2)}, \label{eq3.4}
\end{equation}
where the operator basis satisfies,
\begin{eqnarray}
X_{0}^{(1)} &=& \overline{\rho}_{1}\otimes\overline{\rho}_{3},\qquad
X_{0}^{(2)}  =  \overline{\rho}_{2}\otimes\overline{\rho}_{4}, \nonumber \\
X_{1}^{(1)} &=& \sigma^{z}_{1}\otimes\overline{\rho}_{3},\qquad
X_{1}^{(2)}  =  \sigma^{z}_{2}\otimes\overline{\rho}_{4}, \nonumber \\
X_{2}^{(1)} &=& {\sigma}^{-}_{1}\otimes{\sigma}^{+}_{3} -
{\sigma}^{+}_{1}\otimes{\sigma}^{-}_{3},
\nonumber \\
X_{2}^{(2)} &=& {\sigma}^{-}_{2}\otimes{\sigma}^{+}_{4} -
{\sigma}^{+}_{2}\otimes{\sigma}^{-}_{4},
\nonumber \\
X_{3}^{(1)} &=& {\sigma}^{-}_{1}\otimes{\sigma}^{+}_{3} +
{\sigma}^{+}_{1}\otimes{\sigma}^{-}_{3},
\nonumber \\
X_{3}^{(2)} &=& {\sigma}^{-}_{2}\otimes{\sigma}^{+}_{4} +
{\sigma}^{+}_{2}\otimes{\sigma}^{-}_{4},
\nonumber \\
X_{4}^{(1)} &=& \overline{\rho}_{1}\otimes{\sigma}^{z}_{3},\qquad
X_{4}^{(2)}  =  \overline{\rho}_{2}\otimes{\sigma}^{z}_{4}, \label{eq3.5} \\
X_{5}^{(1)} &=& \sigma^{+}_{1}\otimes\overline{\rho}_{3},\qquad
X_{5}^{(2)}  =  \sigma^{+}_{2}\otimes\overline{\rho}_{4}, \nonumber \\
X_{7}^{(1)} &=& \sigma^{-}_{1}\otimes\overline{\rho}_{3},\qquad
X_{7}^{(2)}  =  \sigma^{-}_{2}\otimes\overline{\rho}_{4}, \nonumber \\
X_{6}^{(1)} &=&
\left[\frac{\overline{n}+1}{2\overline{n}+1}|0\rangle_{1}\langle
0|-\frac{\overline{n}}{2\overline{n}+1}|1\rangle_{1}\langle
1|\right]\otimes{\sigma}^{+}_{3},
\nonumber \\
X_{6}^{(2)} &=&
\left[\frac{\overline{n}+1}{2\overline{n}+1}|0\rangle_{2}\langle
0|-\frac{\overline{n}}{2\overline{n}+1}|1\rangle_{2}\langle
1|\right]\otimes{\sigma}^{+}_{4},
\nonumber \\
X_{8}^{(1)} &=&
\left[\frac{\overline{n}+1}{2\overline{n}+1}|0\rangle_{1}\langle
0|-\frac{\overline{n}}{2\overline{n}+1}|1\rangle_{1}\langle
1|\right]\otimes{\sigma}^{-}_{3},
\nonumber \\
X_{8}^{(2)} &=&
\left[\frac{\overline{n}+1}{2\overline{n}+1}|0\rangle_{2}\langle
0|-\frac{\overline{n}}{2\overline{n}+1}|1\rangle_{2}\langle
1|\right]\otimes{\sigma}^{-}_{4}.  \nonumber
\end{eqnarray}
Clearly, in the representation (\ref{eq3.4}) above, the dynamical
equations (\ref{eq2.4}) decouple in systems (1) and (2) and we
describe the effect of the generator of motion ${\cal{L}}^{(k)}$ on
the systems $k=1,2,$ as
\begin{eqnarray}
{\cal{L}}^{(k)}X_{0}^{(k)} &=& 0, \nonumber \\
{\cal{L}}^{(k)}X_{1}^{(k)} &=& -i\alpha_{k}X_{2}^{(k)}, \nonumber \\
{\cal{L}}^{(k)}X_{2}^{(k)} &=&
-2i\alpha_{k}(X_{1}^{(k)}-X_{4}^{(k)})-\gamma_{\text{eff}}X_{2}^{(k)}+i\Delta_{k}X_{3}^{(k)},
\nonumber \\
{\cal{L}}^{(k)}X_{3}^{(k)} &=& i\Delta_{k}X_{2}^{(k)}
-\gamma_{\text{eff}}X_{3}^{(k)},\nonumber \\
{\cal{L}}^{(k)}X_{4}^{(k)} &=&
i\alpha_{k}X_{2}^{(k)}-2\gamma_{\text{eff}}X_{4}^{(k)}, \label{eq3.6} \\
{\cal{L}}^{(k)}X_{5}^{(k)} &=&
-i\omega_{k}X_{5}^{(k)}-i\alpha_{k}X_{6}^{(k)}, \nonumber \\
{\cal{L}}^{(k)}X_{7}^{(k)} &=&
i\omega_{k}X_{7}^{(k)}+i\alpha_{k}X_{8}^{(k)}, \nonumber \\
{\cal{L}}^{(k)}X_{6}^{(k)} &=&
-i\alpha_{k}X_{5}^{(k)}-(\gamma_{\text{eff}}+i\omega_{k+2})X_{6}^{(k)}, \nonumber \\
{\cal{L}}^{(k)}X_{8}^{(k)} &=& i\alpha_{k}X_{7}^{(k)}-
(\gamma_{\text{eff}}-i\omega_{k+2})X_{8}^{(k)}. \nonumber
\end{eqnarray}
Here, we have introduced the effective decoherence rate
$\gamma_{\text{eff}}\equiv(2\overline{n}+1)\gamma$ as well as the
detuning $\Delta_{k}=\omega_{k}-\omega_{k+2}$. The matrix
representation of the generator ${\cal{L}}^{(k)}$ in the operator
basis $\left\{X_{m}^{(k)}\right\}$ is thus
\begin{widetext}
\begin{equation}
{\cal{L}}^{(k)} = \left[
\begin{array}{c|cccc|cc|cc} 0 & & &
& & & &
& \\
\hline & 0 & -i2\alpha_{k} & 0 & 0 & & & & \\
& -i\alpha_{k} & -\gamma_{\text{eff}} & i\Delta_{k} & i\alpha_{k}
& & & & \\
& 0 & i\Delta_{k} & -\gamma_{\text{eff}} & 0 & & & & \\
& 0 & 2i\alpha_{k} & 0 & -2\gamma_{\text{eff}} & & & & \\
\hline & & & & & -i\omega_{k} & -i\alpha_{k} & & \\
& & & & & -i\alpha_{k} & -(\gamma_{\text{eff}}+i\omega_{k+2}) & & \\
\hline & & & & & & & i\omega_{k} & i\alpha_{k} \\
& & & & & & & i\alpha_{k} & -(\gamma_{\text{eff}}-i\omega_{k+2})
\end{array}\right]. \label{eq3.7}
\end{equation}
The dynamical equation of the density operator (\ref{eq3.4}) follows
from (\ref{eq2.4}):
\begin{eqnarray}
\frac{d}{dt}\rho_{1234}(t)&=&\sum_{m,n=0}^{8}\left[\frac{d}{dt}c_{m}^{(1)}(t)\right]
c_{n}^{(2)}(t)X_{m}^{(1)}\otimes X_{n}^{(2)} + c_{m}^{(1)}(t)
\left[\frac{d}{dt}c_{n}^{(2)}(t)\right] X_{m}^{(1)}\otimes
X_{n}^{(2)}\nonumber \\
&=& \sum_{m,n=0}^{8}\left[{\cal{L}}^{(1)}
c_{m}^{(1)}(t)X_{m}^{(1)}\right]\otimes c_{n}^{(2)}(t)X_{n}^{(2)} +
c_{m}^{(1)}(t)X_{m}^{(1)}\otimes\left[{\cal{L}}^{(2)}
c_{n}^{(2)}(t)X_{n}^{(2)}\right], \label{eq3.7c}
\end{eqnarray}
and it decouples for $k=1$ and $k=2$, so expressed as a vector, the
coefficients
\begin{equation}
{\bf{c}}^{(k)}(t)=\left[c_{0}^{(k)}(t),\ldots,c_{m}^{(k)}(t),\cdots,c_{8}^{(k)}(t)\right]^{\text{T}}
\label{eq3.7a}
\end{equation}
are determined by the differential equation
\begin{equation}
\frac{d}{dt}{\bf{c}}^{(k)}(t)={\cal{L}}^{(k)}{\bf{c}}^{(k)}(t),
\label{eq3.7b}
\end{equation}
\end{widetext}
where ${\cal{L}}^{(k)}$ is given in (\ref{eq3.7}). The relevant part
of the equation of motion of system $(k)$ is represented in the
subspace spanned by $X_{0}^{(k)},X_{1}^{(k)},X_{5}^{(k)},$ and
$X_{7}^{{(k)}}$, i.e.\ by the dynamical evolution of the
coefficients $c_{0}^{(k)}(t),c_{1}^{(k)}(t),c_{5}^{(k)}(t),$ and
$c_{7}^{(k)}(t)$. Consequently, the projection operators
${\cal{P}}^{(k)}$ and ${\cal{Q}}^{(k)}$ are defined as
\begin{equation}
{\cal{P}}^{(k)} = \left[
\begin{array}{c|cccc|cc|cc}
1& & & & & & & & \\
\hline
 & 1 & 0 & 0 & 0 & & & & \\
 & 0 & 0 & 0 & 0 & & & & \\
 & 0 & 0 & 0 & 0 & & & & \\
 & 0 & 0 & 0 & 0 & & & & \\
 \hline
 &   &   &   &   & 1 & 0 & &  \\
 &   &   &   &   & 0 & 0 & &  \\
\hline
 &   &   &   &   &  &  &1 & 0  \\
 &   &   &   &   &  &  &0 & 0
\end{array}
\right], \label{eq3.8}
\end{equation}
and
\begin{equation}
{\cal{Q}}^{(k)} = \left[
\begin{array}{c|cccc|cc|cc}
 0 & & & & & & & & \\
\hline
 & 0 & 0 & 0 & 0 & & & & \\
 & 0 & 1 & 0 & 0 & & & & \\
 & 0 & 0 & 1 & 0 & & & & \\
 & 0 & 0 & 0 & 1 & & & & \\
 \hline
 &   &   &   &   & 0 & 0 & &  \\
 &   &   &   &   & 0 & 1 & &  \\
\hline
 &   &   &   &   &  &  &0 & 0  \\
 &   &   &   &   &  &  &0 & 1
\end{array}
\right], \label{eq3.9}
\end{equation}
where $(k)$ indicate that the projection operators operate
exclusively on the subsystems $(k)$. The equation of motion of the
relevant density operator
\begin{equation}
\tilde{\rho}_{1234}(t)={\cal{P}}^{(1)}{\cal{P}}^{(2)}\rho_{1234}(t)
=\text{Tr}_{34}[\rho_{1234}(t)]
\otimes\overline{\rho}_{3}\otimes\overline{\rho}_{4}, \label{eq3.9a}
\end{equation}
therefore describes the subspace in which the mini-reservoirs
represented by atoms $3$ and $4$ remain in the thermal reference
state $\overline{\rho}$ at all times. Note, however, that the
dynamical evolution of the coefficients representing the relevant
part of the dynamics is influenced by the dynamical evolution of
coefficients that represent the irrelevant part of the dynamics.
This ensures that memory effects of the reservoir are imposed on the
dynamics of the relevant subsystem, i.e.\ the reservoir feeds back
into the subsystem.

\begin{figure}[!ptb]
\centerline{\includegraphics[width=8.6cm]{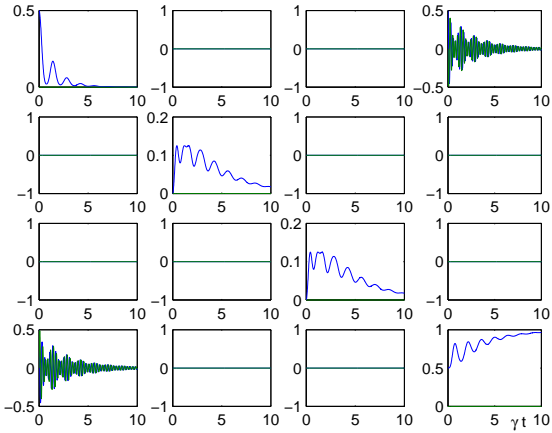}} \caption{Time
evolution of the density matrix elements for the parameters
$\Delta_{1}=\Delta_{2}=2$, $\alpha_{1}=\alpha_{2}=2$, $\gamma=0.5$
and ${\overline{n}}=0$. The frequency $\omega_{1}$ is assumed to be
$10$. The structure of the density matrix remains  of the same form
given in Eq.\ (\ref{eq4.6}) for all parameters used in this paper. }
\label{Fig4.0}
\end{figure}

The initial state of the total system $\rho_{1234}(t=0)$ is given as
\begin{equation}
\rho_{1234}(t=0) =
\rho_{12}(t=0)\otimes\overline{\rho}_{3}\otimes\overline{\rho}_{4},
\label{eq3.10}
\end{equation}
where, $\rho_{12}(t=0)$ is defined in (\ref{eq3.2}). Since the
differential equation of motion is linear, we can solve it for each
tensor product term in (\ref{eq3.2}) independently, and add up the
solutions. For the first tensor product contribution
$|1\rangle_{1}\langle 1|\otimes|1\rangle_{2}\langle 1|$ of
$\rho_{12}(t=0)$ we arrive at the vector of coefficients
\begin{equation}
{\bf{c}}^{(1)}(t=0) =
\left[1,\frac{\overline{n}+1}{2\overline{n}+1},0,\cdots,0\right]^{\text{T}}
={\bf{c}}^{(2)}(t=0). \label{eq3.11a}
\end{equation}
For the second contribution $|0\rangle_{1}\langle
0|\otimes|0\rangle_{2}\langle 0|$, we obtain the following initial
condition
\begin{equation}
{\bf{d}}^{(1)}(t=0) =
\left[1,\frac{-\overline{n}}{2\overline{n}+1},0,\cdots,0\right]^{\text{T}}=
{\bf{d}}^{(2)}(t=0), \label{eq3.12a}
\end{equation}
the third contribution $|0\rangle_{1}\langle
1|\otimes|0\rangle_{2}\langle 1|$ leads to the following
coefficients
\begin{equation}
{\bf{f}}^{(1)}(t=0) = \left[0,0,0,0,0,0,0,1,0\right]^{\text{T}}=
{\bf{f}}^{(2)}(t=0), \label{eq3.13a}
\end{equation}
while for the last contribution $|1\rangle_{1}\langle
0|\otimes|1\rangle_{2}\langle 0|$ we obtain
\begin{equation}
{\bf{g}}^{(1)}(t=0) = \left[0,0,0,0,0,1,0,0,0\right]^{\text{T}}=
{\bf{g}}^{(2)}(t=0). \label{eq3.14b}
\end{equation}
The time-evolution of the relevant density operator in the
appropriate subspace projected to
$\overline{\rho}_{3}\otimes\overline{\rho}_{4}$ is described by the
dynamical evolution of the projected vectors of coefficients
${\cal{P}}^{(k)}{\bf{c}}^{(k)}(t),{\cal{P}}^{(k)}{\bf{d}}^{(k)}(t),{\cal{P}}^{(k)}{\bf{f}}^{(k)}(t),$
and ${\cal{P}}^{(k)}{\bf{g}}^{(k)}(t)$, where $k=1,2$. We calculate
the time evolution of the vectors of coefficients
${\bf{c}}^{(k)}(t),{\bf{d}}^{(k)}(t),{\bf{f}}^{(k)}(t),$ and
${\bf{g}}^{(k)}(t)$ numerically from the equation of motion
(\ref{eq3.7b}), with the initial conditions given in
(\ref{eq3.11a})-(\ref{eq3.14b}). The time evolution of the relevant
density operator (\ref{eq3.9a}) is then given by
\begin{widetext}
\begin{eqnarray}
\tilde{\rho}_{1234}(t)&=&
\frac{1}{2}\sum_{m=c,d,f,g}\left[\begin{array}{cc}
\frac{\overline{n}}{2\overline{n}+1} m^{(1)}_{0}(t)+ m^{(1)}_{1}(t)
&
m^{(1)}_{5}(t) \\
m^{(1)}_{7}(t) &
\frac{\overline{n}+1}{2\overline{n}+1}m^{(1)}_{0}(t)-m^{(1)}_{1}(t)
\end{array}
\right] \nonumber \\
&& \otimes \left[\begin{array}{cc}
\frac{\overline{n}}{2\overline{n}+1} m^{(2)}_{0}(t)+ m^{(2)}_{1}(t)
&
m^{(2)}_{5}(t) \\
m^{(2)}_{7}(t) &
\frac{\overline{n}+1}{2\overline{n}+1}m^{(2)}_{0}(t)-m^{(2)}_{1}(t)
\end{array}
\right] \otimes\overline{\rho}_{3}\otimes\overline{\rho}_{4}.
\label{eq3.15}
\end{eqnarray}
\end{widetext}
This describes the time evolution of  the density operator
(\ref{eq3.4}) projected onto the subspace where the atoms 3 and 4
remain in the thermal state (\ref{eq2.1}) at all times. The
evolution of the reduced systems operator $\rho_{12}(t)$ of the
subsystem $1-2$ follows from Eq.\ (\ref{eq3.15}) by tracing out the
near reservoir systems $3$ and $4$. The dynamical evolution of the
reduced density operator
$\rho_{12}(t)\equiv\text{Tr}_{34}[\tilde{\rho}_{1234}(t)]$ in matrix
representation of the computational basis is displayed in Fig.\
\ref{Fig4.0} for the initial condition given in (\ref{eq3.2}) and
for certain interaction parameters with the near environments. Of
course, $\rho_{12}(t)$ describes a real density matrix that
satisfies the positivity condition since the map from which it has
been derived is completely positive.

\section{Dynamics of entanglement quantified by concurrence}\label{Sec.4}
In this section we consider the dynamics of the entanglement of the
bipartite system $1-2$ in the non-Markovian reservoir assuming the
system density operator $\rho_{12}(t=0)$ to be in the state
(\ref{eq3.2}) initially. We take Wootters concurrence
\cite{Wootters1,Wootters2} to quantify the amount of entanglement
present in the subsystem $\rho_{12}(t)$. The concurrence of a
bipartite, arbitrary density operator $\rho$ of two qubits is
defined as
\begin{equation}
{\cal{C}}(\rho) =
\text{max}\left\{0,\lambda_{1}-\lambda_{2}-\lambda_{3}-\lambda_{4}
\right\}, \label{eq4.1}
\end{equation}
where the $\lambda_{i}$ are the square roots of the eigenvalues of
$\rho\tilde{\rho}$ in descending order. Here $\tilde{\rho}$  is the
result of applying the spin-flip operation to the complex
conjugation or transpose of $\rho$:
\begin{equation}
\tilde{\rho} =
(\sigma_{y}\otimes\sigma_{y})\rho^{\ast}(\sigma_{y}\otimes\sigma_{y}),
\label{eq4.2}
\end{equation}
where the complex conjugation is taken in the standard computational
basis. The concurrence is related to the entanglement of formation
${{E}}_{f}(\rho)$ by the following function
\cite{Wootters1,Wootters2}
\begin{equation}
{{E}}_{f}(\rho) ={\cal{E}}[{\cal{C}}(\rho)], \label{eq4.3}
\end{equation}
where,
\begin{equation}
{\cal{E}}[{\cal{C}}(\rho)] =
h\left[\frac{1+\sqrt{1-{\cal{C}}^{2}(\rho)}}{2}\right],
\label{eq4.4}
\end{equation}
and
\begin{equation}
h(x) = -x\log_{2}x-(1-x)\log_{2}(1-x). \label{eq4.5}
\end{equation}

\begin{figure}[!ptb]
\centerline{\includegraphics[width=7cm]{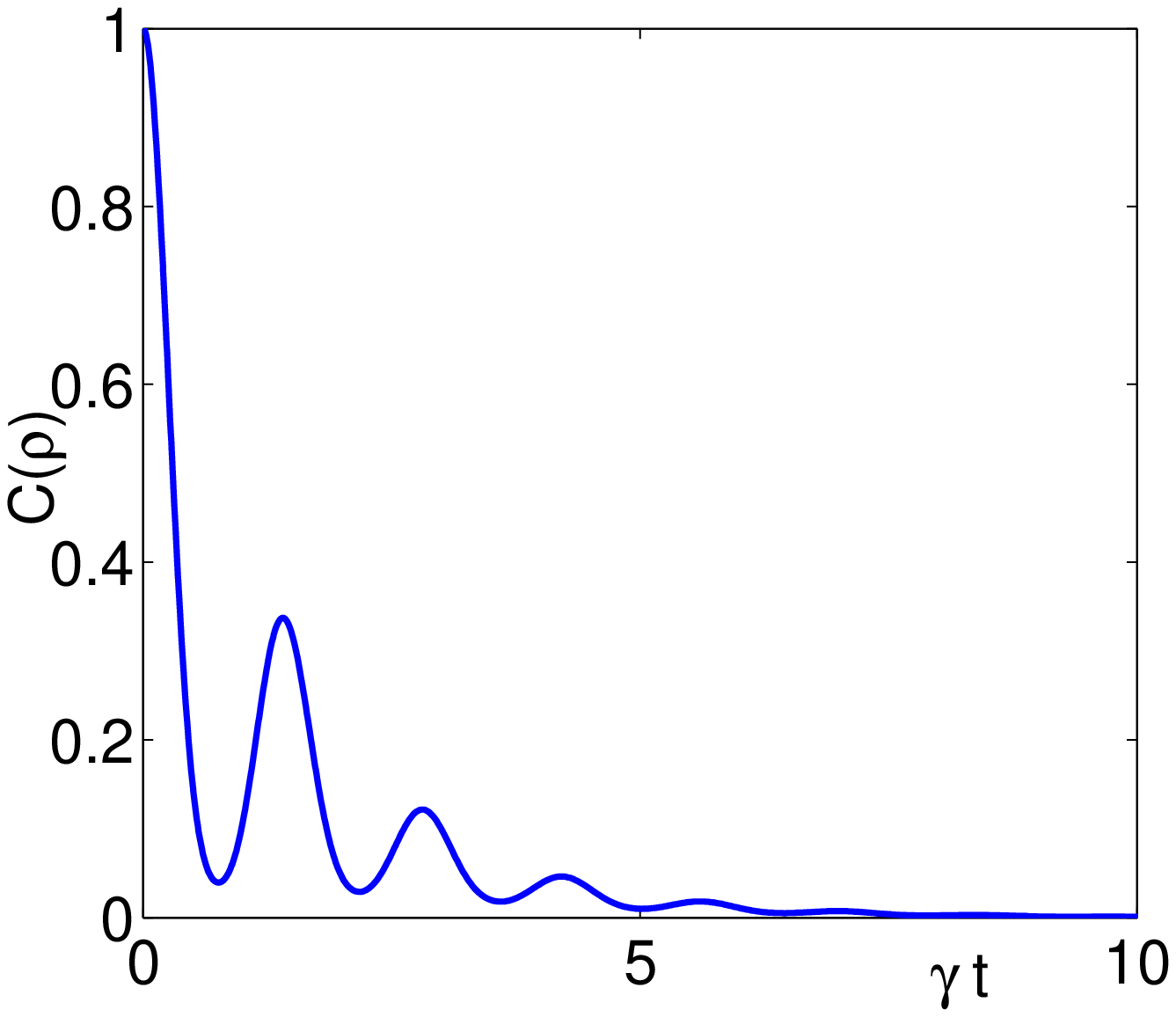}}
\caption{Evolution of the concurrence of an initially maximally
entangled Bell state which decays into the non-Markovian reservoir.
The dimensionless parameters of the reservoir are
$\Delta_{1}=\Delta_{2}=2$, $\alpha_{1}=\alpha_{2}=2$, $\gamma=0.5$
and ${\overline{n}}=0$. The frequency $\omega_{1}$ is assumed to be
$10$. The concurrence displays and oscillating behavior due to the
memory effects of the non-Markovian reservoir. Entanglement which is
initially lost due to decoherence is partially restored by the
memory effects. Because of the large detuning, entanglement does not
completely vanish in a finite time. } \label{Fig4.1}
\end{figure}

\begin{figure}[!ptb]
\centerline{\includegraphics[width=7cm]{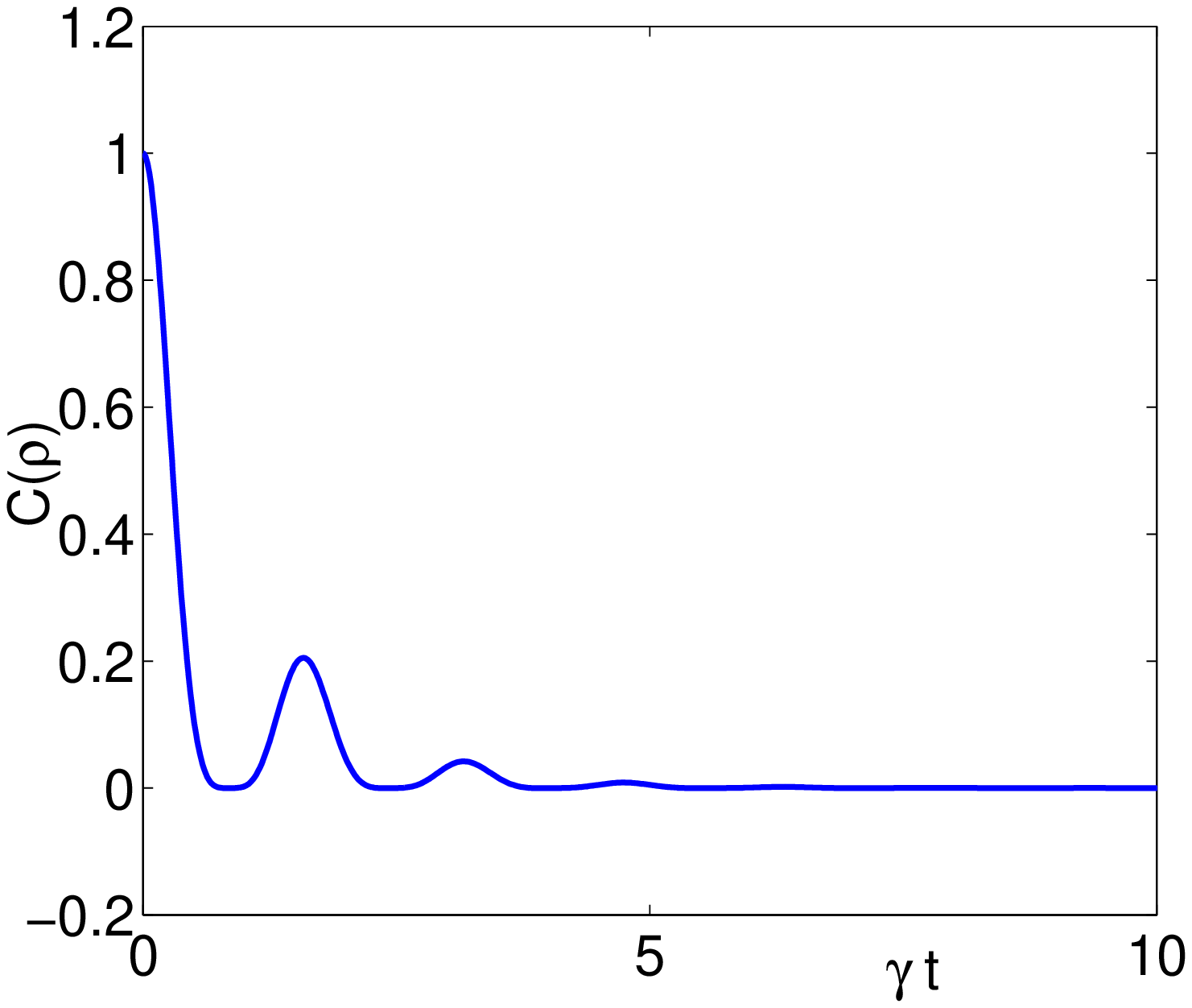}}
\caption{Evolution of the concurrence of an initially maximally
entangled Bell state which decays into the non-Markovian reservoir.
The dimensionless parameters of the reservoir are
$\Delta_{1}=\Delta_{2}=0$, $\alpha_{1}=\alpha_{2}=2$, $\gamma=0.5$
and ${\overline{n}}=0$. The frequency $\omega_{1}$ is assumed to be
$10$. As in Fig.\ \ref{Fig4.1} the concurrence displays and
oscillating behavior due to the memory effects of the non-Markovian
reservoir. However, the entanglement completely vanishes at finite
times before it is partially recovered in contrast to Fig.\
\ref{Fig4.1}. This is a consequence of the vanishing detuning.}
\label{Fig4.2}
\end{figure}

\begin{figure}[!ptb]
\centerline{\includegraphics[width=7cm]{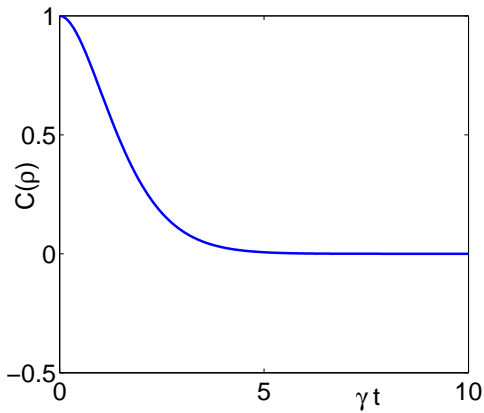}}
\caption{Evolution of the concurrence of an initially maximally
entangled Bell state which decays into the non-Markovian reservoir.
The dimensionless parameters of the reservoir are
$\Delta_{1}=\Delta_{2}=0$, $\alpha_{1}=\alpha_{2}=0.5$, $\gamma=0.5$
and ${\overline{n}}=0$. The frequency $\omega_{1}$ is assumed to be
$10$. The concurrence does not display an oscillating behavior any
more since the interaction with the near environment is comparable
weak and damping dominates. However, the decay of the concurrence is
non-exponential which is a manifestation of the non-Markovian
reservoir. } \label{Fig4.3}
\end{figure}

\begin{figure}[!ptb]
\centerline{\includegraphics[width=7cm]{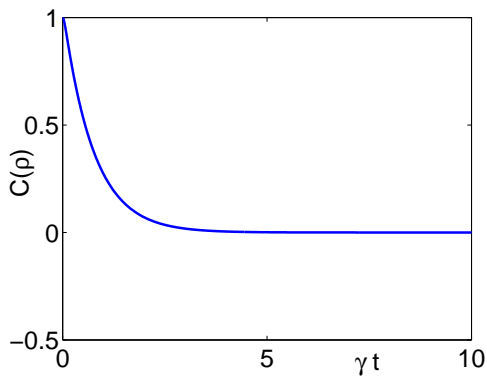}}
\caption{Evolution of the concurrence of an initially maximally
entangled Bell state which decays into a Markovian reservoir which
is achieved when $\alpha/\gamma \to 0$. The dimensionless parameters
of the reservoir are $\Delta_{1}=\Delta_{2}=0$,
$\alpha_{1}=\alpha_{2}=3$, $\gamma=27$ and ${\overline{n}}=0$ giving
an effective Markovian decay rate of $\Gamma=1/3$ (see text). The
frequency $\omega_{1}$ is assumed to be $10$. } \label{Fig4.4}
\end{figure}

The structure of the density matrix of the two-atom system remains
in the following form during the time-evolution
\begin{equation}
\rho_{12}(t) = \left[
\begin{array}{cccc}
a(t) & 0 & 0 & f(t) \\
0 & b(t) & 0 & 0  \\
0 & 0 & c(t) & 0  \\
f^{\ast}(t) & 0 & 0 & d(t)
\end{array}
\right], \label{eq4.6}
\end{equation}
where $a(t)+b(t)+c(t)+d(t)=1$. This property is a consequence of the
special initial condition and the fact that the environment,
regardless of whether it is Markovian or not, can not create
coherence that was not present initially. The time evolution of the
density matrix is shown in Fig.\ \ref{Fig4.0} for a certain set of
parameters. This particular form of the density matrix (\ref{eq4.6})
allows us to analytically express the concurrence as
\begin{equation}
{\cal{C}}[\rho_{12}(t)] = 2\text{max}\{0,|f(t)|-\sqrt{b(t)c(t)}\}.
\label{eq4.7}
\end{equation}
In Figs.\ \ref{Fig4.1}-\ref{Fig4.4} we plot the evolution of the
concurrence for different parameters of the non-Markovian bath. As a
result of the memory effects the concurrence displays an oscillatory
time dependence, especially when the coupling parameter $\alpha$ is
strong. In the case of zero detuning, $\Delta=0$, the entanglement
can vanish after a finite time even when the initial state of the
two-atom system is a maximally entangled Bell state, as shown in
Fig.\ \ref{Fig4.2}. However, due to the memory inherent in the
non-Markovian bath, the entanglement is partially restored. The
two-atom system remembers its initial degree of entanglement to some
extent, which leads to a revival of the entanglement. The collapse
and revival of the entanglement displays a damped oscillatory
behavior in time, whose frequency depends on the coupling parameter
$\alpha$. Thus the coupling parameter $\alpha$ characterizes the
memory ability of the reservoir to some extent. In case when the
detuning $\Delta$ differs from zero, the entanglement does not
completely vanish after a finite time, as displayed in Fig.\
\ref{Fig4.1}. The detuning suppresses the decoherence and thus
extends the decoherence time. This is because the mini-reservoir
atoms `shield' atoms $1$ and $2$ from the reservoir. When the
damping rate $\gamma$ is of the same order or greater than the
coupling parameter $\alpha$, the decay of the entanglement is
damping oriented and oscillations do not appear, as seen in Fig.\
\ref{Fig4.3}. However, the effect of the non-Markovian bath is still
visible, as the concurrence displays a non-exponential decay in
time, and the decoherence time is extended. The last plot Fig.\
\ref{Fig4.4} displays the concurrence in the Markovian limit, which
is achieved when $\alpha/\gamma\to 0$ but $\alpha^{2}/\gamma$
remains finite \cite{Salo}. In this case the Markovian time scale is
given as \cite{Salo}
\begin{equation}
\Gamma=\frac{1}{(2\overline{n}+1)^2}\frac{\alpha^{2}}{\gamma},
\label{eq4.8}
\end{equation}
leading to an effective decoherence rate of $\Gamma=1/3$ for the
parameters used in Fig.\ \ref{Fig4.4}.

\begin{figure}[!ptb]
\centerline{\includegraphics[width=7cm]{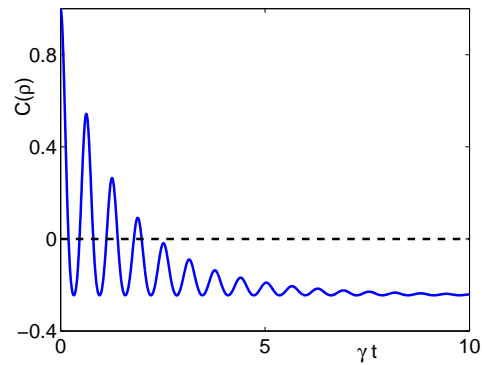}}
\caption{Evolution of the concurrence of an initially maximally
entangled Bell state which decays into the non-Markovian reservoir
of finite temperature with mean photon number $\overline{n}=0.2$.
Here, we do not plot the actual concurrence but
${\cal{C}}[\rho_{12}(t)]=2\{|f|-[b(t)c(t)]^{1/2}\}$ instead of
${\cal{C}}[\rho_{12}(t)]=2\text{max}\{|f|-[b(t)c(t)]^{1/2}\}\geq 0$
to see how negative it becomes. The dimensionless parameters of the
reservoir are $\Delta_{1}=\Delta_{2}=0$, $\alpha_{1}=\alpha_{2}=5$,
$\gamma=1/3$ and ${\overline{n}}=0.2$. The frequency $\omega_{1}$ is
assumed to be $10$. As in Figs.\ \ref{Fig4.1} and \ref{Fig4.2} the
concurrence displays and oscillating behavior due to the memory
effects of the non-Markovian reservoir. In contrast to the zero
temperature case, however, the concurrence vanishes at a finite time
completely and cannot be retrieved. In addition, during the time
period where the concurrence can be partially recovered there are
extended periods of time where it vanishes completely.}
\label{Fig4.5}
\end{figure}

\begin{figure}[!ptb]
\centerline{\includegraphics[width=7cm]{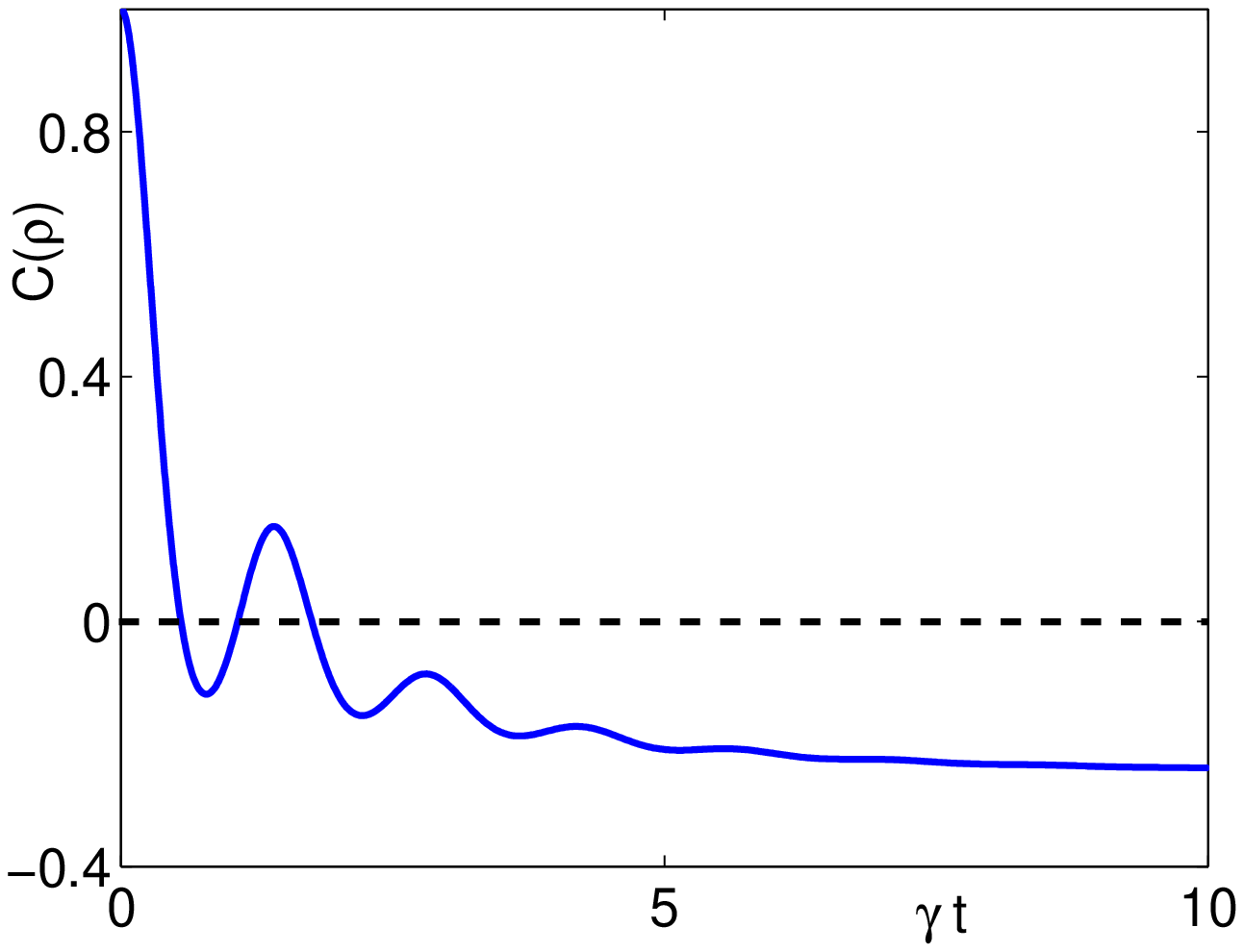}} \caption{Same as
Fig.\ \ref{Fig4.5} but with parameters $\Delta_{1}=\Delta_{2}=2$,
$\alpha_{1}=\alpha_{2}=2$, $\gamma=0.5$ and ${\overline{n}}=0.2$.
The frequency $\omega_{1}$ is assumed to be $10$. } \label{Fig4.6}
\end{figure}

\begin{figure}[!ptb]
\centerline{\includegraphics[width=7cm]{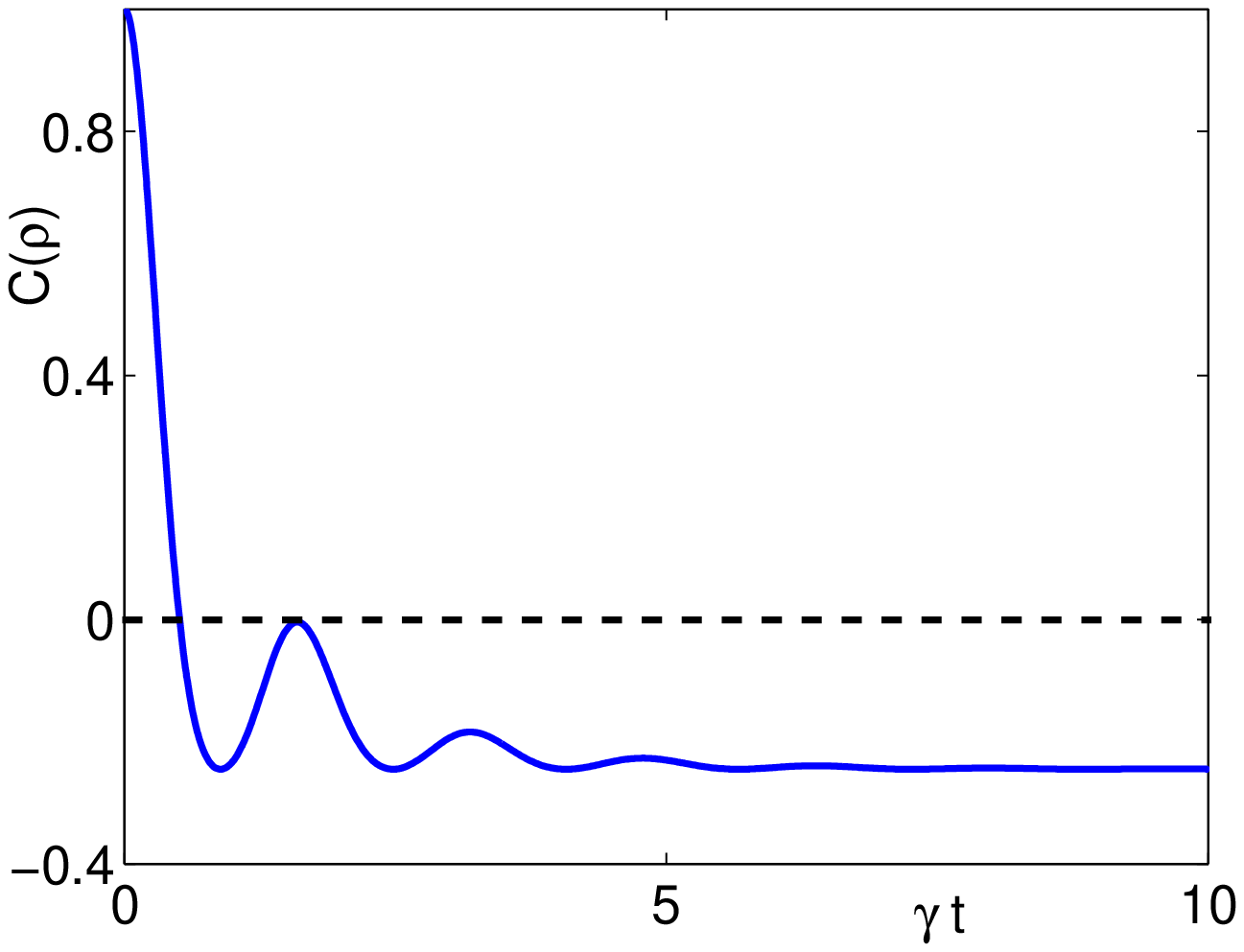}} \caption{Same as
Fig.\ \ref{Fig4.5} but with parameters $\Delta_{1}=\Delta_{2}=0$,
$\alpha_{1}=\alpha_{2}=2$, $\gamma=0.5$ and ${\overline{n}}=0.2$.
The frequency $\omega_{1}$ is assumed to be $10$. } \label{Fig4.7}
\end{figure}

\begin{figure}[!ptb]
\centerline{\includegraphics[width=7cm]{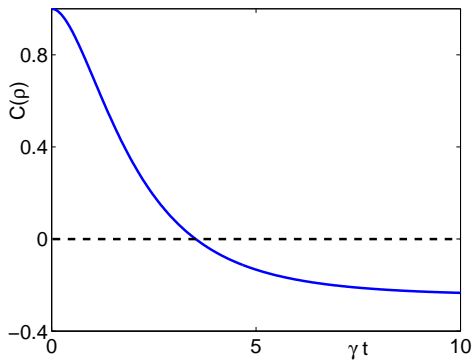}} \caption{Same as
Fig.\ \ref{Fig4.5} but with parameters $\Delta_{1}=\Delta_{2}=0$,
$\alpha_{1}=\alpha_{2}=0.5$, $\gamma=1$ and ${\overline{n}}=0.2$.
The frequency $\omega_{1}$ is assumed to be $10$. This figure
displays the strong damping case in which the effect of the near
environment is comparably weak. However non-Markovian effects are
clearly visible since the decay of the concurrence is
non-exponential. As a result the period of time during which the
concurrence is maintained is extended compared to the Markovian
decay. } \label{Fig4.8}
\end{figure}

\begin{figure}[!ptb]
\centerline{\includegraphics[width=7cm]{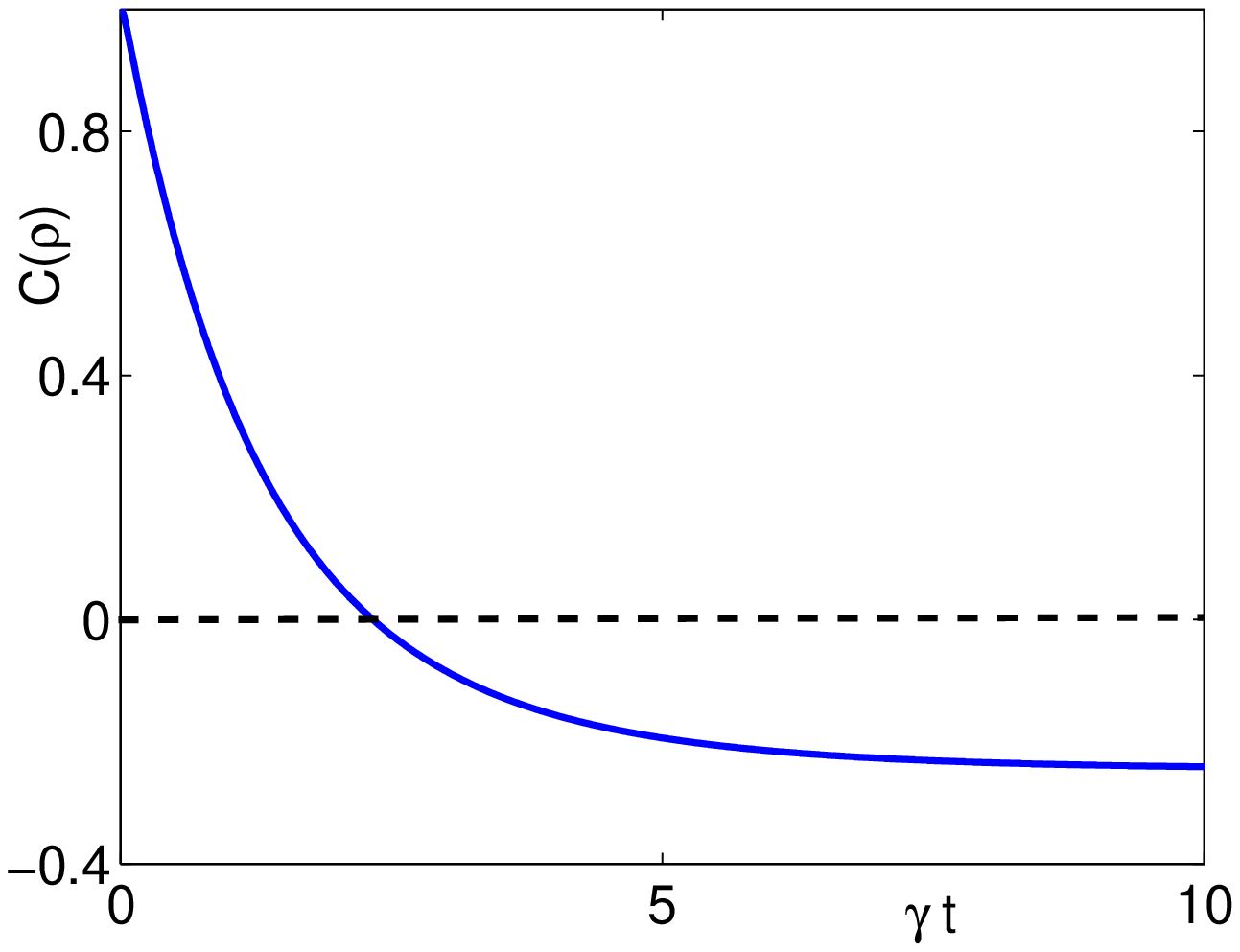}}
\caption{Evolution of the concurrence in the Markovian limit of the
temperature reservoir. The parameters are $\Delta_{1}=\Delta_{2}=0$,
$\alpha_{1}=\alpha_{2}=3$, $\gamma=27$ and ${\overline{n}}=0.2$
leading to an effective Markovian decay rate of $\Gamma\approx 1/6$.
The frequency $\omega_{1}$ is assumed to be $10$. In comparison to
Fig.\ \ref{Fig4.8} the concurrence is maintained during a shorter
period of time.} \label{Fig4.9}
\end{figure}

An important observation with respect to entanglement is that for an
open quantum system, entanglement can die off at finite times. It
has recently been shown that this can happen not only when one
starts with mixtures of entangled quantum states, but also in cases
of initially pure and maximally entangled states, if the entangled
system is coupled to a {\textit{thermal}} reservoir at finite
temperature \cite{Eberly1,Eberly2,Carvalho}. In other words
non-local disentanglement times are in general shorter than local
decoherence times. In practical quantum computation schemes which
operate at finite temperature this is of particular importance since
some quantum computational algorithms rely on entanglement. We
therefore ask next, to what extent a non-Markovian reservoir affects
the decoherence of the entangled system in the case of an
environment with non-zero temperature. This is illustrated in Figs.\
\ref{Fig4.5}-\ref{Fig4.9} for an environment with a mean photon
number of $\overline{n}=0.2$, i.e.\ a hot reservoir.
In order to visualize the periods over which the concurrence becomes
zero more clearly, we do not display the actual concurrence given in
(\ref{eq4.7}) but the quantity $2\{|f(t)|-[b(t)c(t)]^{1/2}\}$, which
is less than zero when the concurrence vanishes. In contrast to the
zero temperature case we can see that the concurrence vanishes
suddenly after a finite time for all plots, a behavior which has
been predicted for Markovian reservoirs in \cite{Eberly1,Eberly2}.
In addition, in cases of strong coupling to the near environments,
when $\alpha>\gamma$, the concurrence can vanish for a period of
time, before being partially recovered due to the memory effect of
the non-Markovian reservoir. As the interaction parameter $\alpha$
with the near environment becomes stronger, these periods of time
when concurrence completely vanishes become shorter, and the rate at
which the concurrence partially revives increases as shown in Figs.\
\ref{Fig4.5}-\ref{Fig4.7}. The time until there is no more revival
of the concurrence becomes longer when there is some detuning
between the system under consideration and the near environment as
can be seen by comparison of Figs.\ \ref{Fig4.6} and \ref{Fig4.7}.
When $\gamma>\alpha$ and the damping dominates, the effect of the
non-Markovian reservoir becomes weaker and the evolution of the
concurrence does not display an oscillating behavior any more.
However the decay of the concurrence is non-exponential and the
period of time before it vanishes completely can be extended in
comparison with a Markovian reservoir, as can be seen by comparing
Fig.\ \ref{Fig4.8} with Fig.\ \ref{Fig4.9}, which shows the
Markovian limit. For the parameters given in Fig.\ \ref{Fig4.9} the
effective Markovian decay rate $\Gamma$ for the Markovian limit can
be calculated from Eq.\ (\ref{eq4.8}) to be $\Gamma\approx 1/6$.
Although this effective decay rate is considerably smaller than the
decay rate $\gamma =1$ used in the non-Markovian reservoir of Fig.\
\ref{Fig4.8}, the period of time before the concurrence suddenly
vanishes is shorter than for the non-Markovian reservoir. Thus a
non-Markovian reservoir helps to maintain concurrence over a larger
period of time in comparison to a Markovian reservoir.
\section{Summary and Conclusions}\label{Sec.Concl}
In this paper we have discussed the effects of a non-Markovian bath
on the decoherence of an entangled two-atom system where the atoms
are assumed to be two-level systems representing qubits. The
non-Markovian bath is described by a model that leads to an exact
Nakajima-Zwanzig type master equation that can be solved without any
approximations, as discussed by Salo et al.\ in \cite{Salo}. The
advantage of this master equation is that it leads to a completely
positive time evolution and therefore non-physical quantum states do
not arise, a situation which can appear in parameterized
phenomenological non-Markovian master equations \cite{Barnett1}.

We discuss the time-evolution of the entanglement of an initially
maximally entangled Bell state in this non-Markovian reservoir. The
entanglement is quantified by the concurrence
\cite{Wootters1,Wootters2} which can be derived in an analytical
form since the density matrix of the two-atom system retains a
certain structure during time evolution. The dynamics of the
concurrence displays some interesting features which are
manifestations of the memory effects inherent in a non-Markovian
reservoir. In particular, the entanglement can vanish after a finite
time but, depending on the parameters of the reservoir, it can be
partially restored. The revival of the concurrence is an effect that
is specific to the non-Markovian reservoir since the reservoir
remembers that the system was initially completely entangled, and
some of this entanglement can be restored in the principle system
even though it becomes zero during the time evolution.

In the case of finite temperature baths it is well known that the
concurrence of an entangled system can experience ``sudden death''
when the system decays into a Markovian reservoir
\cite{Eberly1,Eberly2,Carvalho}. This sudden death of the
concurrence can not be prevented in a thermal non-Markovian
reservoir either. However, due to the memory effect in the
non-Markovian reservoir the concurrence can revive even when it
completely vanished after a finite period of time. Depending on the
parameters used, the concurrence displays a collapse and revival
structure in time, which finally ends in sudden death. The time of
the sudden death can be extended by the non-Markovian reservoir in
comparison with the Markovian reservoir. This is an effect of the
non-exponential decay of the concurrence in a non-Markovian
reservoir compared with its exponential decay in a Markovian
reservoir.

\acknowledgments Helpful discussions with Stefano Bettelli, Momtchil
Peev, Martin Suda, and especially with Janne Salo and Stig Stenholm
are gratefully acknowledged.
\bibliographystyle{unsrt}

\end{document}